\documentclass[prx,superscriptaddress,twocolumn,footinbib]{revtex4-2}
\pdfoutput=1
\usepackage[colorlinks=true,urlcolor=blue]{hyperref}
\usepackage{amsfonts}
\usepackage{graphicx}
\usepackage{placeins}
\usepackage{amsmath}
\usepackage{xcolor}
\usepackage{color}
\usepackage{bm}
\usepackage{ulem}
\usepackage{array}
\usepackage{babel}

\definecolor{red}{rgb}{0.75,0,0}
\definecolor{blue}{rgb}{0,0,1}
\definecolor{green}{rgb}{0,0.5,0}

\newcommand{\red}[1]{{\color{black} #1}}

\begin{document}
	
	\title{Topological defects lead to energy transfer in active nematics}
	
	\author{Daniel J. G. Pearce}
	\affiliation{Department of Theoretical Physics, University of Geneva, 1205 Geneva, Switzerland}
	
	\author{Berta Mart\'inez-Prat}
	\affiliation{Department of Materials Science and Physical Chemistry, Universitat de Barcelona, Barcelona, 08028, Spain}
	\affiliation{Institute of Nanoscience and Nanotechnology, IN2UB, Universitat de Barcelona, Barcelona, 08028, Spain}
	
	\author{Jordi Ign\'{e}s-Mullol}
	\affiliation{Department of Materials Science and Physical Chemistry, Universitat de Barcelona, Barcelona, 08028, Spain}
	\affiliation{Institute of Nanoscience and Nanotechnology, IN2UB, Universitat de Barcelona, Barcelona, 08028, Spain}

	\author{Francesc Sagu\'{e}s}
	\affiliation{Department of Materials Science and Physical Chemistry, Universitat de Barcelona, Barcelona, 08028, Spain}
	\affiliation{Institute of Nanoscience and Nanotechnology, IN2UB, Universitat de Barcelona, Barcelona, 08028, Spain}

 \date{\today}
 
\begin{abstract}
  Active nematics are fluids in which the components have nematic symmetry and are driven out of equilibrium due to the microscopic generation of an active stress. When the active stress is high, it drives flows in the nematic and can lead to the proliferation of topological defects, a state we refer to as defect chaos. Using numerical simulations of active nematics at low Reynolds number, we observe energy transfer from long to short length scales during defect chaos. We demonstrate that this energy transfer is driven by the exchange between variations in the orientation and degree of order in the nematic that predominantly occurs during defect creation and annihilation. We then show that the primary features of energy transfer during defect chaos scale with the active length scale. Finally, we identify a second regime that forms extended bend walls instead of point like topological defects, which we refer to as bend wall chaos. The bend walls grow and extend over the whole system and result in an energy transfer from short to long length scales.
\end{abstract}

	\maketitle

	Active matter is the study of systems that are driven out of equilibrium by the generation of stress at the component scale~\cite{ramaswamy2010mechanics,ramaswamy2017active,marchetti2013hydrodynamics, sagues23}. Examples of active materials range from the micrometer scale, for example swimming sperm, through to kilometer scale shoals of fish~\cite{pearce2018cellular,schoeller2020collective,makris2009critical}. Active nematics are active fluids in which the components have nematic symmetry~\cite{doostmohammadi2018active}. This describes a large class of biosystems including bacterial colonies, the cytoskeleton and epithelial sheets~\cite{juelicher2007active,balasubramaniam2022active,saw2018biological,mueller2019emergence,yashunsky2024topological,you2018geometry}. Among the most striking examples of active nematics are those made from reconstituted cytoskeletal components, namely microtubules and kinesins, an experimental system which has been used to demonstrate many properties of active suspensions~\cite{sanchez2012spontaneous,guillamat2018active,pearce2021orientational,decamp2015orientational}. One of the most prominent features of active nematics is the presence of topological defects, point like discontinuities in the orientation of the nematic components characterized by a half integer winding number~\cite{giomi2015geometry}. In addition, nematic topological defects have an orientation which effects their motion relative to other defects in the material~\cite{pearce2021properties,tang2017orientation,vromans2016orientational}. Topological defects in active nematics also lead to distinct local force patterns, in particular the polar symmetry of $+1/2$ defects leads them to self propelling~\cite{thampi2014instabilities,giomi2014defect,giomi2015geometry,pearce2021orientational}. When the active stress is sufficiently high, it generates large flows which destabilize the nematic orientation, leading to defect proliferation and small scale vortices~\cite{giomi2015geometry}, a state we refer to as defect chaos. 

    Defect chaos is visually similar to classical turbulence, with the distinction that it occurs at low Reynolds number, when classical turbulence is impossible~\cite{kundu2015fluid}. This similarity has driven a lot of interest in understanding the mechanism behind this phenomenon~\cite{giomi2015geometry,koch2021role,opathalage2019self,doostmohammadi2017onset}. Giomi showed that the vortices generated during defect chaos have exponentially distributed areas, and identified the scaling parameter as the ``active length scale'', the scale at which active and elastic forces balance~\cite{giomi2015geometry}; these predictions have been experimentally corroborated~\cite{Guillamat17}. The active length scale, often denoted $l_\alpha$, is the length scale at which deformations in the nematic director field become unstable due to the flows generated by the active stress, and has been shown to be one of the primary length scales in active nematics, dictating features in the flow field and director field, as well as defect densities and spatial correlations~\cite{giomi2015geometry,pearce2021orientational,hemingway2016correlation}. Giomi also showed that power law scaling was expected in the kinetic energy spectrum in the high wave number range. Simulations by Alert et al.~\cite{alert2020universal} reported additional universal scaling of flow spectra in the range of small wave numbers, with a separating crossover dictated by the active length scale. A more general form of scaling was confirmed in the microtubule/kinesin experimental system taking into account external dissipation \cite{martinez2021scaling}.  
    Classical turbulence is also characterized by an energy cascade, in which kinetic energy is transferred between length scales~\cite{kundu2015fluid}. Alert et al. \cite{alert2020universal,alert2022active} also addressed this question, considering both elastic and kinetic contributions in an active nematic that does not allow for the presence of topological defects. They concluded that, within the energy scaling regime, there is no energy transfer. Results by Urzay et al.~\cite{urzay2017multi} and Carenza et al.~\cite{carenza2020cascade}, considering purely kinetic contributions, are essentially consistent with the lack of energy cascades, although authors in the latter work observed some residual form of energy transfer due to elastic stresses at intermediate length scales. However, in the microtubule/kinesin experimental system evidence for energy transfer between scales has been observed, \red{for example in Fig.4.2 of}~\cite{martinez2022active}, which leaves this question somewhat unresolved.
    
    
	In this manuscript, we numerically study chaotic flows in active nematics at low Reynolds number. We consider the total energy stored in the system, that being the sum of the kinetic energy and the Landau De-Gennes free energy. We derive the dissipative terms and observe a clear energy transfer between length scales. We demonstrate that this energy transfer is due to an exchange of energy between variations in the orientation of the nematic and its scalar order parameter, a process which happens predominantly during defect annihilation and creation. We confirm that, in a broad area of parameter space, the active length scale remains the dominant one. Finally, we identify a new regime in the parameter space that is characterized by the presence of bend walls instead of topological defects. This change is associated with a transition from viscous to elastic dissipation.

	We study an incompressible two-dimensional active nematic. The nematic texture is described by the $\bm{Q}$ tensor which combines the nematic orientation $\bm{n}$ and the local degree of nematic order $S\in[0,1]$ into a single tensorial order parameter $Q_{ij} = S(n_i n_j - \delta_{ij}/2)$. The dynamics of the velocity $\bm{v}$ and the nematic tensor $\bm{Q}$ are described by the respective equations~\cite{beris1994thermodynamics}:
	\begin{align}
    \label{eq:mom}
		\rho D_t \bm{v} &= \eta \nabla^2\bm{v} + \nabla.(\bm{\sigma}^p + \bm{\sigma}^E + \alpha \bm{Q}), \qquad \nabla.\bm{v} = 0,\\
		D_t\bm{Q} &= \lambda S \bm{u} + \frac{1}{\gamma}\bm{H} + \bm{Q}.\bm{\omega} - \bm{\omega}.\bm{Q}.
	\end{align}
	As commonly written, $\bm{u}$ and $\bm{\omega}$ respectively represent  the strain rate and vorticity tensors, $\alpha$ is the activity, $\lambda$ is the dimensionless flow aligning parameter, $\rho$ is a constant density, $\eta$ is the shear viscosity, and $\gamma$ is the rotational viscosity. The passive stress is given by $\bm{\sigma}^p = -P\mathbb{I}-\lambda S\bm{H} + \bm{Q}.\bm{H} - \bm{H}.\bm{Q}$, Here $P$ is the pressure which is fixed by the incompressibility condition. $\bm{\sigma}^E$ is the traceless symmetric Ericksen stress, see \red{Eq.~S13 in}~\cite{SI}. Finally, the molecular tensor $\bm{H} = -\delta F^{LDG}/\delta \bm{Q}$; here $F^{LDG}$ is the Landau De-Gennes free energy given by
	\begin{equation}
		F^{LDG} = \frac{K}{2}\int \Big[|\nabla\bm{Q}|^2 + \frac{1}{\epsilon^2}\textrm{tr}\bm{Q}^2(\textrm{tr}\bm{Q}^2 - 1)\Big] \textrm{d}A.
	\end{equation}
	In the expression above, $K$ denotes the single elastic constant and $\epsilon$ is a length scale which dictates the typical length scale over which $S$ varies, often referred to as the defect core radius. 
    
\begin{figure}[t]
		\centering
		\includegraphics[width=\columnwidth]{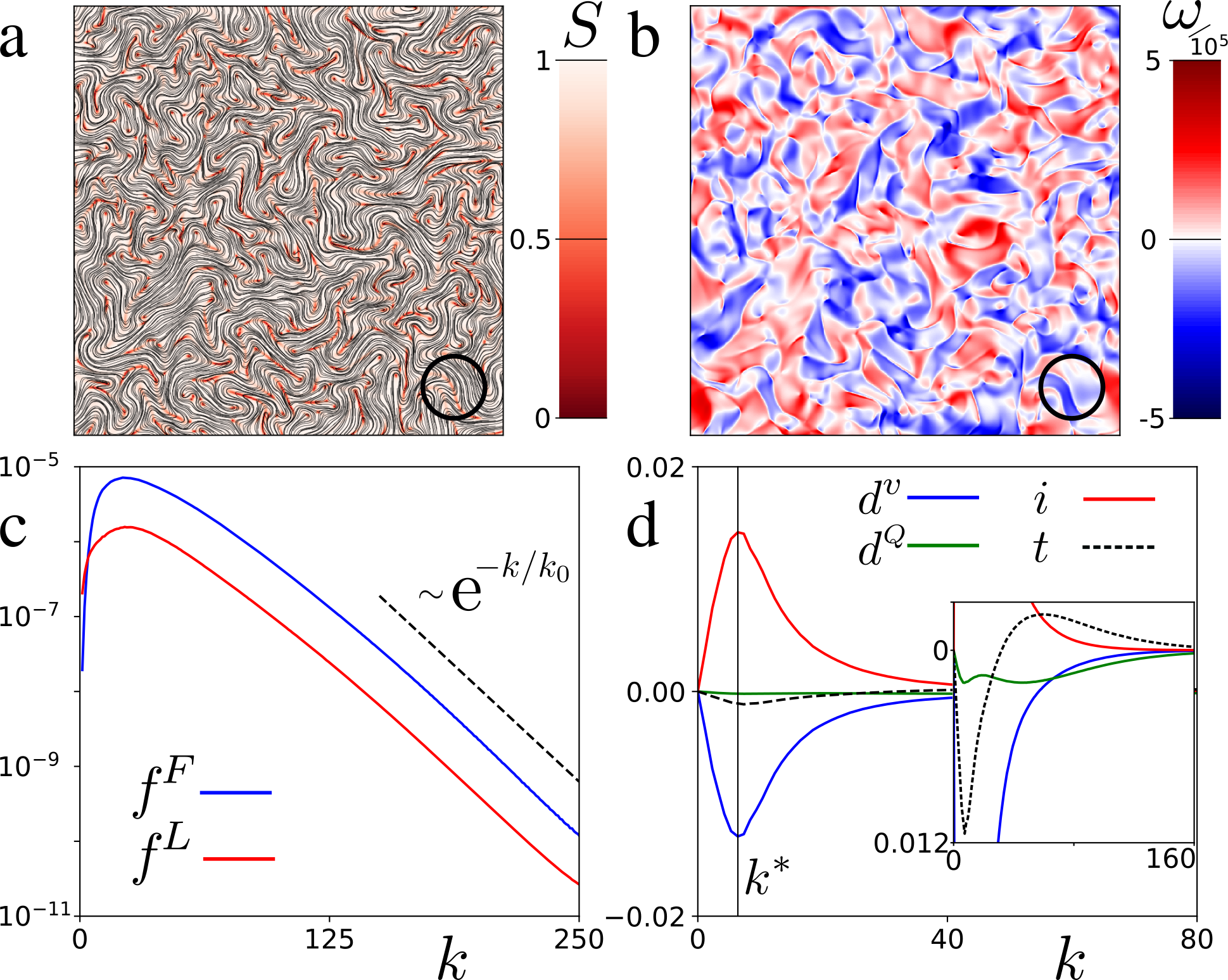}
		\caption{\label{fig:f1} Defect chaos. (a\&b) \red{Snapshot of a} simulated active nematic showing the director field (lines) and scalar order parameter (color) (a) and vorticity (b). (c) \red{Time-averaged} spectra for the  Landau De-Gennes free energy showing the distortion (blue) and scalar order parameter (red) contributions separately. The dashed line shows an exponential distribution with $k_0 \sim 17.5$; this scale is predominantly governed by the defect core radius, $\epsilon$, rather than the activity, see \red{section III of}~\cite{SI}. (d) \red{Time-averaged} dissipation spectra for a simulated active nematic with total energy transfer given by the black dashed line. The wave number associated with the peak energy injection \red{($i$)} is identified as $k^{*}\approx 7.5$, which also corresponds to the peak in stored elastic energy; \red{see Fig.~S2 in~\cite{SI}}. The length scale associated with this wave number is indicated by the black circle in (a)\&(b). (d inset) Magnified image of (d) showing better the features of $d^Q$ and $t$. }
    \end{figure}

    We simulate the active nematic on a $512\times512$ periodic square grid to reach a steady state of defect chaos, see Fig.~\ref{fig:f1}a,b and SI Movie 1. We set the length of the simulation area to be $L=1$ and the single elastic constant to be $K=1$, which define our length and energy scales respectively. We set the rotational viscosity $\gamma = 1$ and re-scale time by $\tau = \gamma L^2/K$. All results are given in re-scaled units. We set $\epsilon = 1.15/512$, which fixes the defect scale to be of the order of the grid spacing, and $\alpha = -0.125\times512^2$, which results in a simulation containing of the order of $10^2$ defects reflecting typical experimental data sets~\cite{pearce2021orientational}. We enforce low Reynolds number by simulating Stokes flow, setting the left hand side of Eq.~\ref{eq:mom} to zero; we set the viscosity to $\eta=0.25$. We set flow alignment $\lambda=0$, we discuss the effect of flow alignment in more detail in the \red{section V of}~\cite{SI}. Similar parameter values have been shown to reproduce well the behavior of microtubule based active nematics~\cite{pearce2021orientational}. All simulations are initialized with a random orientation field which is then pre-simulated for $10^5$ simulation steps; results are then collected over the following $10^6$ simulation steps. \red{The spectra shown in figures are calculated by taking the average of the instantaneous spectra obtained every $10^3$ simulation steps throughout the data collection period. }

	The total energy of an active nematic is the sum of the Landau De-Gennes free energy, $F^{LDG}$, and the kinetic energy given by $F^{KE} = \frac{1}{2}\int\rho\bm{v}^2\textrm{d}A$. We observe the expected kinetic energy spectra in both the high and low wavenumbers, see \red{section IV of}~\cite{SI}. Here we note that the Landau De-Gennes free energy is itself the sum of two components, $F^{LDG} = F^F + F^L$. $F^F = \frac{K}{2}\int|\nabla\bm{Q}|^2\textrm{d}A$ describes the energetic cost of variations in $\bm{Q}$; in the nematic limit ($S=1$) this becomes identical to the Frank free energy. $F^L = \frac{K}{2\epsilon^2}\int\textrm{tr}\bm{Q}^2(\textrm{tr}\bm{Q}^2 - 1)\textrm{d}A$ contains Landau terms governing the isotropic to nematic transition; this penalizes deviations of the scalar order parameter from $S=1$. This is a key difference with previous work which was performed at the $S=1$ limit, such that $F^L=0$~\cite{alert2020universal,alert2022active}. The contributions to the Landau De-Gennes free energy over Fourier space are plotted in Fig.~\ref{fig:f1}c (here we use $f^F$ and $f^L$ to denote the energy densities associated with $F^F$ and $F^L$, respectively). We observe a well defined peak in the elastic energy spectrum; it has been shown that this peak is associated with the active length scale, $l_\alpha = \sqrt{K/|\alpha|}$~\cite{alert2020universal,alert2022active} and \red{section III of}~\cite{SI}. We also see a clear exponential distribution of both $f^F$ and $f^L$, indicating the existence of another length scale in the system related to wavenumber $k_0$; we confirm that this is primarily governed by the defect core radius, $\epsilon$ see \red{section III of}~\cite{SI}.

	When studying energy transfer, we consider the rate of change of total energy in the active nematic, namely $\dot{F} = \dot{F}^{KE} + \dot{F}^{LDG}$. This is a key distinction to previous works which consider only dissipation of kinetic energy in keeping with studies on classical turbulence~\cite{urzay2017multi,carenza2020cascade}. 
    The rate of change of total energy is calculated as:
	\begin{alignat}{2}
        \label{eq_fdot}
		\dot{F} &= D^v + D^Q + I,\\
		D^v &= -2\eta\int  u_{ij} u_{ij}\ \textrm{d}A &&= \int d^v(k)\ \textrm{d}k,\\
		D^Q &= -\frac{1}{\gamma}\int H_{ij}H_{ij}\ \textrm{d}A &&= \int d^Q(k)\ \textrm{d}k,\\
		I &= -\alpha\int  u_{ij}Q_{ij}\ \textrm{d}A &&= \int i(k)\ \textrm{d}k.
	\end{alignat}
	Here we have used Einstein notation and repeated indexes are summed over. We identify energy dissipation due to shear viscosity and relaxation of the $Q$ tensor with $D^v$ and $D^Q$, respectively. We identify $I$ with injection of energy due to active stresses, $I$. We use lower case to indicate the dissipation at a specific wavenumber $k$ in Fourier space, see \red{section I.A of}~\cite{SI} for a derivation.
	
	When the system is at a steady state, such as that of fully developed defect chaos, there is no net gain or loss of energy, $\langle\dot{F}\rangle=0$. \red{This is true for all wave numbers, hence $\dot{f}(k)=0$. If the energy injection and dissipation at a given wave number do not exactly balance, the excess (or deficit) energy is transferred to (or from) other length scales. This allows us to define the energy transfer $t(k)$ according to
    \begin{equation}
        \dot{f}(k) = d^v(k) + d^Q(k) + i(k) + t(k).
    \end{equation}
    } 
    \red{Thus $t(k)$ negative (positive) implies that energy is transferred from (to) a certain wave number.} The contributions to $\dot{f}$ are plotted in Fourier space for an active nematic undergoing defect chaos in Fig.~\ref{fig:f1}d. The largest magnitude factors are active injection, $i$, and shear dissipation, $d^v$, which are peaked around a single length scale. This is in agreement with previous research which has identified the active length scale, $l_\alpha$, as dominant during chaotic active nematic flows~\cite{giomi2015geometry,hemingway2016correlation}. Energy dissipation due to relaxation of the Landau De-Gennes free energy is comparatively much smaller.

    We directly assess energy transfer between length scales by calculating \red{$t = -d^v - d^Q - i$}, plotted as a black dashed line in Fig.~\ref{fig:f1}d. \red{The system has an excess influx of energy around the peak of injection and shear dissipation located at $k^*$. This excess is transferred away to other wavenumbers, indicated by $t(k)<0$ around $k^*$. Energy is transferred in to higher wavenumbers where $t(k)>0$, where it is then dissipated.} This indicates there is an energy transfer from large to small length scales in the dynamics of the active nematic at low Reynolds number, contrary to typical Stokes flow and previous studies on defect-free active nematics~\cite{alert2020universal,alert2022active,kundu2015fluid}.

    The energy transfer between scales we observe is different from classical turbulence, for a few key reasons. First, Stokes flow omits any non-linear terms from the momentum equation and precludes the possibility of kinetic energy exchange between length scales~\cite{kundu2015fluid}. We confirm that in our simulated active nematic there is no kinetic energy exchange between length scales, specifically $t^{KE}(k) = 0$ for all wave numbers $k$, see Fig.~S1~\cite{SI}; this is in agreement with previous studies that report no kinetic energy transfer in low Reynolds number active nematics~\cite{urzay2017multi,carenza2020cascade}. This implies that any energy transfer between scales must arise from the dissipation of elastic energy.
    
    We now look closely at the dissipation of elastic energy, which is described by the following equations, 
	\begin{alignat}{2}
		\dot{F}^{LDG} &= D^Q - E^f - E^{E} - E^{a},\\
        E^f &= -\lambda \int S H_{ij}u_{ij}\ \textrm{d}A &&= \int e^f(k)\ \textrm{d}k,\\
        E^{E} &= -\int \sigma^E_{ij}u_{ij}\ \textrm{d}A &&= \int e^{E}(k)\ \textrm{d}k,\\
        E^{a} &= \int \omega_{ij}\sigma^a_{ij}\ \textrm{d}A &&= \int e^{a}(k)\ \textrm{d}k.
	\end{alignat}
    Here $\bm{\sigma}^a = \bm{Q}.\bm{H} - \bm{H}.\bm{Q}$ is the antisymmetric part of the passive stress. Here $E^f$, $E^{E}$ and $E^{a}$ represent the exchange of kinetic energy with elastic energy through flow alignment, relaxation of the Ericksen stress and rotation of the director field, respectively; see \red{section I of}~\cite{SI} for a derivation. \red{We adopt the convention here to refer to flux of energy between scales as transfer and a flux of energy between different forms as exchange.} Since we are neglecting flow alignment, $E^f$ is not plotted here. Recent data-based studies on microtubule based active nematics have shown that flow alignment plays a significant role in the dynamics of the director field~\cite{joshi2022data,golden2023physically}; we demonstrate that introducing flow alignment does not significantly change our findings on energy transfer, see \red{section V of}~\cite{SI}. 

    %

    Additionally, we expand upon how energy is dissipated through relaxation of the director field. As in experiments, our simulations allow for variations in the scalar order parameter, which are penalized by the Landau terms in the free energy, denoted $F^L$. This is in addition to energy stored as gradients in the orientation of the director field, denoted $F^F$. Thus there are two ways in which the director field can be relaxed, and we can write $D^Q$ as the sum of three terms
	\begin{alignat}{2}
		D^Q &= D^F + D^L + D^{FL},\\
		D^F &= -\frac{K^2}{\gamma}\int [\partial_k^2Q_{ij}][\partial_l^2Q_{ij}]\ \textrm{d}A &&= \int d^F(k)\ \textrm{d}k,\\
		D^L &= -\frac{K^2}{2\gamma\epsilon^4}\int [S-S^3]^2\ \textrm{d}A &&= \int d^L(k)\ \textrm{d}k,\\
		D^{FL} &= -\frac{2K^2}{\gamma\epsilon^2}\int [1-S^2]Q_{ij}\partial_k^2Q_{ij}\ \textrm{d}A &&= \int d^{FL}(k)\ \textrm{d}k.
	\end{alignat}
	$D^F$ describes energy dissipated by relaxing gradients of $Q$, governed by $F^F$, and $D^L$ describes energy dissipated by relaxing deviations from $S=1$, governed by $F^L$. Finally, $D^{FL}$ describes energy exchange between $f^F$ and $f^L$, see \red{section I.D of}~\cite{SI} for a full derivation.

    We can now write the full dissipation of elastic energy as
    \begin{equation}
        \dot{F}^{LDG} = D^F + D^L + D^{FL} - E^f - E^{E} - E^{a}.
    \end{equation}
    \red{As before, at steady state there is no build up of elastic energy at any given wave number, $\dot{f}^{LDG}(k)=0$. Thus any discrepancy between injection and dissipation of elastic energy at a given wave number indicates energy transfer between scales. This allows us to define the transfer of elastic energy between scales according to
    \begin{multline}
        \dot{f}^{LDG}(k) = d^F(k) + d^L(k) + d^{FL}(k) - e^f(k)\\ - e^{E}(k) - e^{a}(k) + t^{LDG}(k).
    \end{multline}
    }
    
\begin{figure}[h]
        \centering
        \includegraphics[width=\columnwidth]{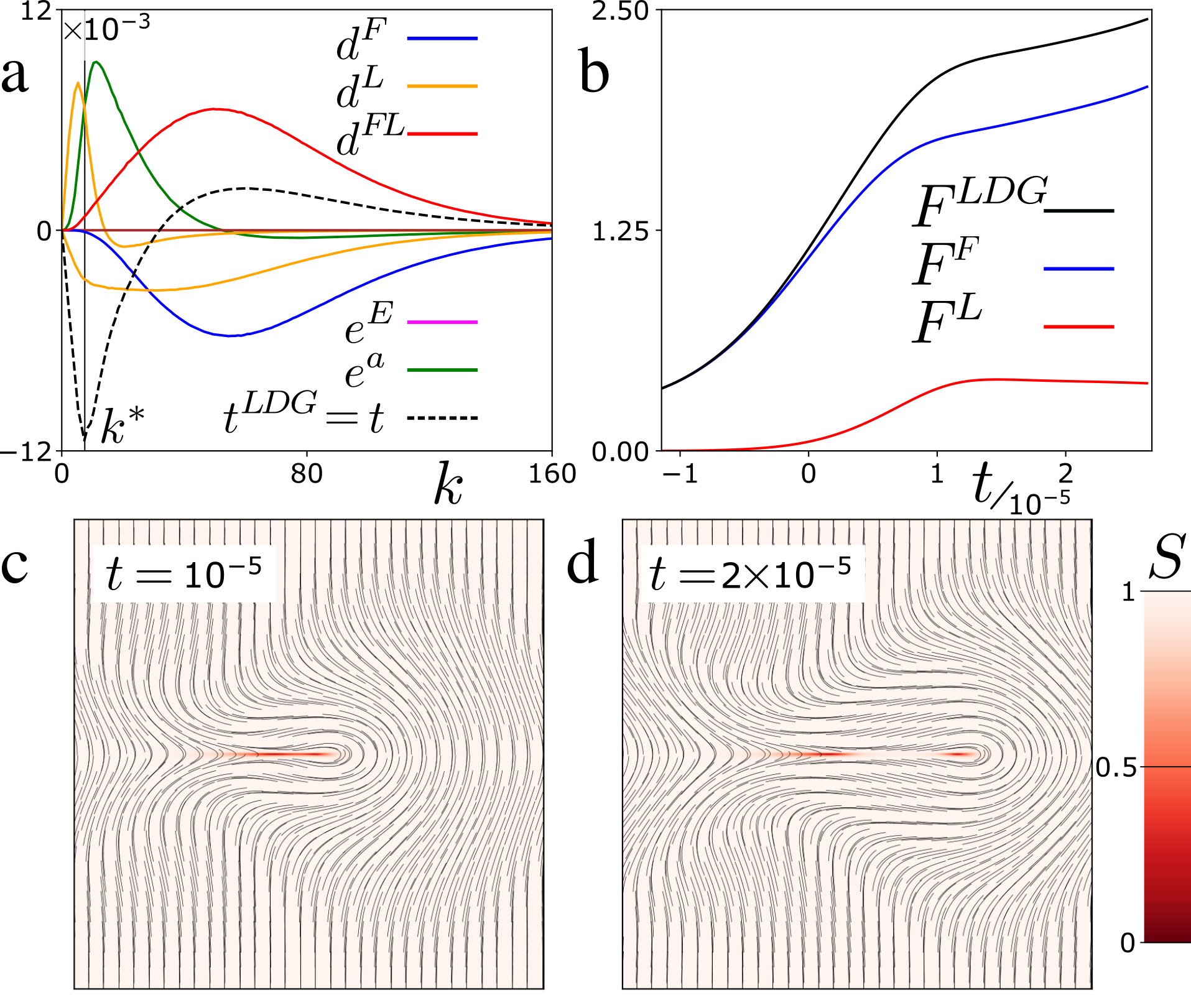}
        \caption{\label{fig:f2} (a) \red{Time-averaged } spectra of the components of the dissipation of elastic energy. The vertical line indicates the previously identified peak in energy injection, $k^*$. \red{Flow alignment, $e^f(k)$, is not plotted here as it is zero.} (b) Total Landau De-Gennes free energy and its components during defect nucleation via bend instability. The difference between the two lines shows the energy stored as deviations from $S=1$. (c\&d) Director field (lines) and scalar order parameter (color) in the time following defect creation via bend instability.}
    \end{figure}

	The contributions to $\dot{f}^{LDG}$ are plotted as a function of wavenumber in Fig.~\ref{fig:f2}a. \red{The combination of the dissipative and exchange terms leads to a non zero energy transfer given by $t^{LDG}(k)$. Since there is no kinetic energy transfer between scales, this accounts for all energy transfer within the system, i.e. $t(k) = t^{LDG}(k)$.} Kinetic energy is exchanged into elastic energy around $k^*$ through a combination of rotation of the director, $e^a$, and Ericksen stress, $e^E$. The latter also exchanges elastic energy back to kinetic energy at shorter wavelengths, resulting in $E^E\approx0$. Thus rotation of the director is the dominant exchange of kinetic to elastic energy. Elastic energy is then dissipated as relaxation of the the Landau De-Gennes free energy \red{at shorter wavelengths}.  
    

    \red{Relaxation of the Landau De-Gennes free energy primarily occurs in two ways, relaxation of $F^F$, given by $d^F$, and relaxation of $F^L$, given by $d^L$. As expected both of these are negative, however they are peaked at different wavenumbers.} The term $d^{FL}$ is positive and represents the exchange of elastic energy between $F^F$ and $F^L$. The primary mechanism through which energy can be exchanged between $F^F$ and $F^L$ is through the co-localization of gradients of $Q$ and deviations from $S=1$. A primary example of this is during the creation of topological defects. Fig.~\ref{fig:f2}b shows the distribution of the elastic energy throughout this process. Defects are created when a constant $Q$ field is no longer stable to perturbations due to the effect of active stress, which occurs at a minimum wavelength that scales with the active length, $l_\alpha$ ~\cite{giomi2015geometry}. In an extensile active nematic, the bend deformation is unstable, as shown in Fig.~\ref{fig:f2}c. The deformation is penalized by the $F^F$ terms in the energy which impose a cost on gradients in the orientation field. These gradients can be reduced by decreasing the order parameter from $S=1$, which is in turn penalized by the $F^L$ terms in the energy. This leads to a co-localization of distortions in the director field with deviations from $S=1$, see Fig.~\ref{fig:f2}c. As the bend distortion grows, the local energy density increases until it becomes favorable to nucleate a pair of topological defects. This results in a pair of discontinuities in the orientation of the director, which is offset by a localized region with $S\approx0$. This process leads to a significant increase in $F^L$ but an overall reduction in the increase of $F^{LDG}$, see Fig.~\ref{fig:f2}b. At this point, the bend distortion is replaced by a pair of topological defects and the region between them now features a stable splay distortion, see Fig.~\ref{fig:f2}d. This process focuses all distortion in the director and the scalar order parameter into a small region localized around the topological defects; thus transferring energy from a long wavelength structure to a short wavelength one.

    Taken together, these observations give us a clear picture of the way energy is injected, transferred and dissipated within active nematics. Energy is injected as kinetic energy, driven by the active stress at the active length scale. There is no kinetic energy cascade in these low Reynolds number flows, and kinetic energy is dissipated either directly, through viscosity ($D^v$), or is converted into elastic energy through either flow alignment ($E^f$), or rotation of the director field ($E^{a}$). Energy converted through flow alignment increases or decreases the order parameter depending on the sign of $\lambda$ leading to an increase in $F^L$. This is because the active stress is aligned with the director field, see \red{section V of}~\cite{SI} for a discussion. \red{Kinetic energy converted through rotation of the director field ($E^{a}$) does not affect the order parameter. Thus $E^{a}$ leads to an increase in elastic energy stored in the form of gradients in the orientation field ($F^F$) around the active length scale.} \red{Long wavelength gradients in the orientation field are unstable and grow in magnitude.} When the gradient in the orientation field becomes high, the local order parameter drops to compensate
    . The decrease in order parameter is highly localized, due to the finite value of $\epsilon$, causing gradients in the orientation field to become highly localized \red{leading to a reduction in wavelength.} \red{A pair of topological defects are generated when the gradient gets focused into a discontinuity, associated with high wavenumber modes.} Thus through the creation of defects, elastic energy is transferred to higher wavenumbers than that at which it is injected. The Ericksen stress further compounds this effect as it exchanges kinetic energy to elastic energy at small wavenumbers, and back at large wavenumbers.

    

    These results are in direct contrast with previous studies on energy transfer in active nematics which were performed at the $S=1$ limit, thus precluding the existence of topological defects and the outlined mechanism~\cite{alert2020universal,alert2022active}. Since topological defects are such a predominant feature in many experimental examples of active nematics, we expect energy transfer to exist in experimentally realizable active nematic systems.

    \begin{figure}[htb]
		\centering
		\includegraphics[width=\columnwidth]{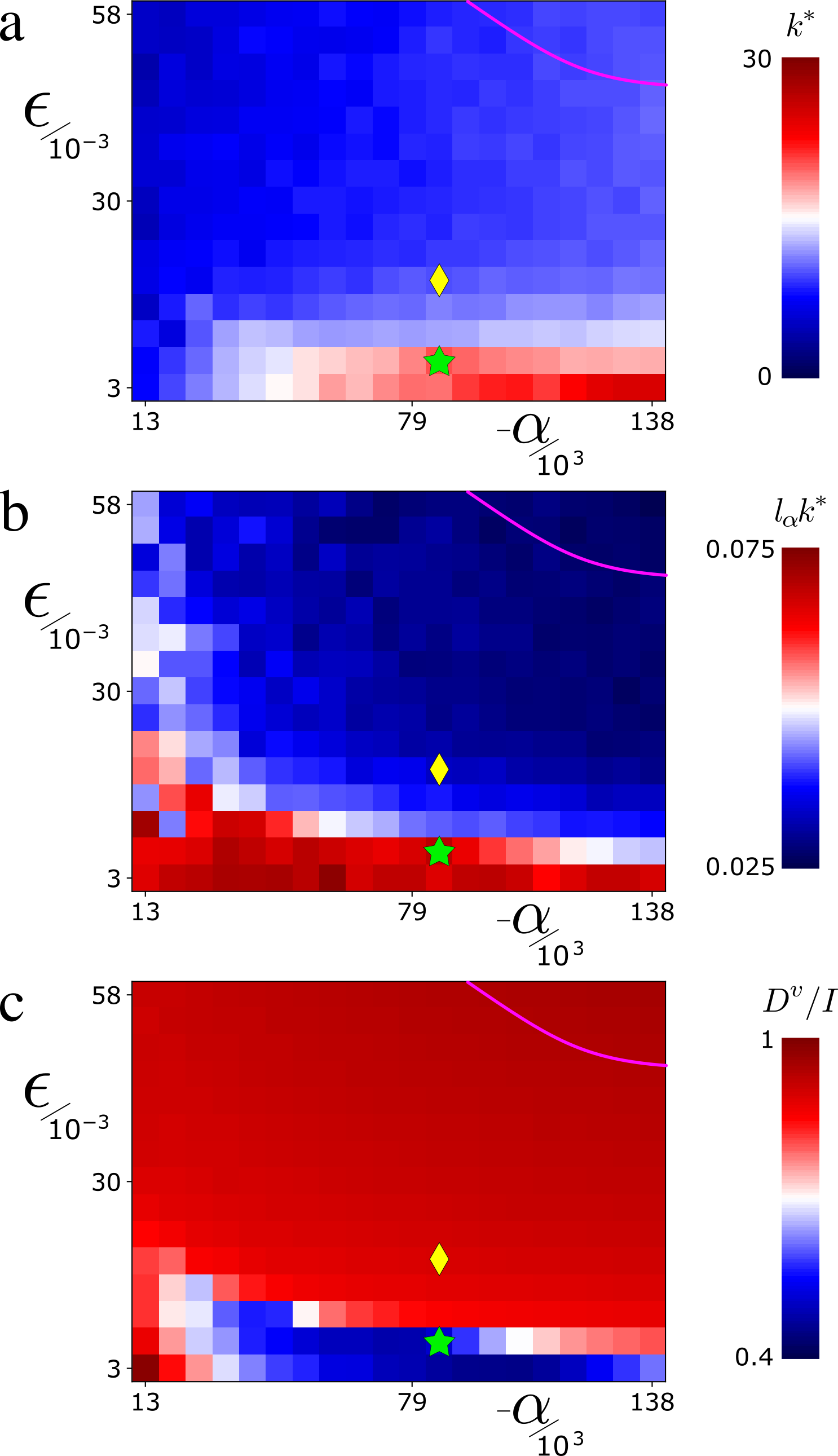}
		\caption{\label{fig:f3} (a) Peak wavenumber, $k^*$, of the energy transfer spectra as a function of activity, $\alpha$, and defect core radius, $\epsilon$. (b) Same as (a) with $k^*$ scaled by the active length scale. (c) Fraction of energy dissipated through viscosity as a function $\alpha$ and $\epsilon$. The yellow diamonds indicate the location of simulations shown in Figs.~\ref{fig:f1},\ref{fig:f2}. The green stars indicate the location of simulations shown in Fig.~\ref{fig:f4}. The pink lines indicate the ratio between defect density and defect size observed in microtubule based experiments, see \red{section VI of}~\cite{SI} for details.  }
    \end{figure}
 
	The active length, $l_\alpha$, governs the scale of many features of an active nematic and can be controlled by varying the activity, $\alpha$. The other important length is the defect core radius, $\epsilon$, which controls the balance between the two terms in the Landau De-Gennes free energy. Fig.~\ref{fig:f3}a shows how the wavelength of the negative valley in the energy transfer, $k^*$, varies with these two length scales. We normalize this wavenumber by the active length scale by plotting $l_\alpha k^*$ which reveals two distinct regimes in the behavior of the active nematic, see Fig.~\ref{fig:f3}b. The parameters resulting in the energy transfer presented in Figs.~\ref{fig:f1}\&~\ref{fig:f2} are represented by the yellow diamond, which is in the regime we recognize as typical defect chaos. By studying experimentally realized microtubule based active nematics, we can approximate the inter defect spacing by $L/\sqrt{N}$; here $N$ is the number of defects observed in an $L^2$ area, see SI Movie 3. We can also estimate $\epsilon$ by fitting to the Landau De-Gennes free energy, see \red{section VI of}~\cite{SI} for details. We obtain a dimensionless ratio between the two of $L/\epsilon\sqrt{N} = 8.66$ which we can use to estimate where in the parameter space we expect typical experiments to lie, indicated by the pink line Fig.~\ref{fig:f3}b. This confirms that experimental active nematics lie in a similar state of defect chaos. See Fig.~S6 of~\cite{SI} for the computed energy transfer obtained from simulations for a point on the pink line.

    \begin{figure}[h]
		\centering
		\includegraphics[width=\columnwidth]{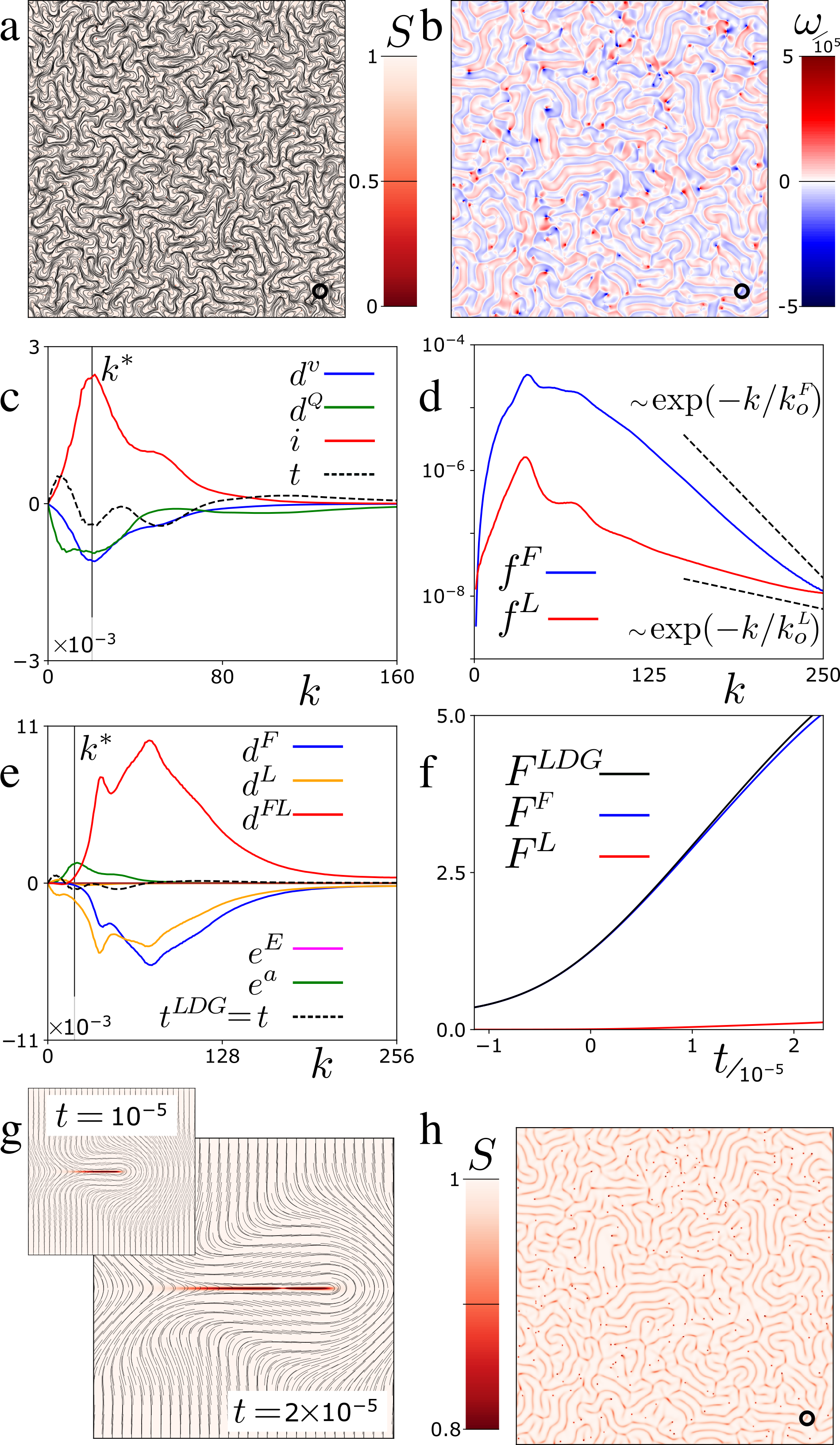}
		\caption{\label{fig:f4} Bend wall chaos. (a) Director field and scalar order parameter and (b) vorticity. The black circle indicates the length scale of energy injection. (c) \red{Time averaged} total energy dissipation spectra showing energy transfer across scales. (d) \red{Time averaged} spectra for the Landau De-Gennes free energy showing the distortion (blue) and order (red) separately. The black dashed lines show the emergence of two exponential distributions. (e) \red{Time averaged} elastic energy dissipation spectra. (f) Energy of the different components of the Landau De-Gennes free energy during bend instability. (g) Bend instability leads to the generation of bend walls which extend over large scales. All results in this figure are simulated with $\alpha = -0.325\times512^2$, $\epsilon = 0.35/512$ indicated by the green star in Fig.~\ref{fig:f3}b.}
    \end{figure}

    To understand the two regimes identified in Fig.~\ref{fig:f3}b, we look at the dominant method of energy dissipation over the same parameter range. The total energy injection rate is given by $I$, which is dissipated by either viscosity through $D^v$ and relaxation of the director $D^Q$. Thus $D^v/I$ gives the fraction of energy dissipated through viscosity, shown in Fig.~\ref{fig:f3}c. Here we see that in typical defect chaos, the dominant method of energy dissipation is viscosity. Defect chaos emerges when $l_\alpha$ is small, hence when activity, $\alpha$, is high. High activity increases the flow speed, and thus strain rate, leading to an increase in $D^v$. Conversely, $D^Q$ does not directly scale with the activity thus during defect chaos $D^v$ dominates energy dissipation. This is similar to the high activity regime identified in defect free active nematics~\cite{alert2022active}. However, we see that for small values of $\epsilon$ the scaling with $l_\alpha$ breaks down and the dominant method of dissipation is $D^Q$, see Figs.~\ref{fig:f3}b,c. When $\epsilon$ is very small, $D^L$ increases and eventually causes $D^Q$ to become the dominant method of energy dissipation, Fig.~\ref{fig:f3}c. In this state, the Landau terms in the free energy dominate and the cost of topological defects becomes very high. Furthermore, when $\alpha$ is small the active length scale is large and fewer defects are expected. Thus we arrive at a state with a low number of defects, see Fig.~\ref{fig:f4} and SI Movie 2. In this regime, the active nematic is still highly dynamic, demonstrating chaotic flows and the same scaling of the kinetic energy spectra, see \red{Fig.~S3 of}~\cite{SI}. This makes it distinct from previously identified low defect states that can display long range order or oscillatory behavior~\cite{oza2016antipolar,oza2016generalized}.
     
	The state identified in Fig.~\ref{fig:f4} occurs when $\epsilon$ is small and variations in the scalar order parameter from $S=1$ are heavily penalized; this is reminiscent of recent numerical studies on active nematics in which the order parameter $S=1$ is fixed and no energy transfer was observed~\cite{alert2020universal,alert2022active}. However, contrary to those studies, the dissipation spectra in this regime display a clear sign of energy transfer between length scales, see Fig.~\ref{fig:f4}c. In this regime, we observe a large energy sink at small wavenumber, driven by $d^Q$, associated with a peak in the energy transfer, $t$. This implies that energy transfer in this regime is from small to large scale, the opposite to that observed in defect chaos. As before, the elastic energy spectra both feature a single peak with exponentially distributed tails, however they now feature different exponents, see Fig.~\ref{fig:f4}d. For further discussion, see \red{section III of}~\cite{SI}. 

    Similar to defect chaos, this new regime features no kinetic energy transfer between length scales, see Fig.~S1~\cite{SI}. Thus all energy transfer between scales occurs as elastic energy. Once again, we study the individual sources and sinks of elastic energy in the system, see Fig.~\ref{fig:f4}e. Similar to defect chaos, the dominant source of elastic energy is rotation of the director, given by $e^a$, which leads to an increase in $F^F$. Again, high wave number distortions in the orientation field co-localize with deviations from $S=1$ resulting in an energy exchange between the two, represented by $d^{FL}$. However, in this regime $d^L$ dominates at small wave numbers, indicating long wavelength dissipation due to relaxation of the order parameter, see Fig.~\ref{fig:f4}e. 

	To examine how energy is transferred to longer wavelengths in this regime, we look again at the behavior of the bend instability, see Fig.~\ref{fig:f4}f. In this regime, the elastic energy increases at a rapid rate throughout the bend instability, without the decrease associated with defect nucleation. The initial bend instability grows in length rather than nucleating topological defects, see Fig.~\ref{fig:f4}g. This is because deviations from $S=1$ have an increased energetic cost, which is incompatible with topological defects which require $S=0$ at their core where the orientation is discontinuous. The energy of the bend distortion can still be reduced by locally decreasing $S$. However, since $\epsilon$ is small, this is associated with a high energetic cost. The result is a slight reduction of $S$ in a narrow region where the distortions in the director can be focused into a bend wall, see Fig.~\ref{fig:f4}g. The bend walls continue to grow, leading to long wavelength structures in the order parameter, which results in the more negative $d^L$ at small wavenumbers. 
    
    The bend walls eventually fill the whole space subdividing it into channels, see Fig.~\ref{fig:f4}h. Locally, a bend wall is translationally invariant along its length, thus the active flow it generates is laminar. Due to the nature of the bend instability, adjacent bend walls are anti-parallel, resulting in the laminar flow channels shown in Fig.~\ref{fig:f4}b. The laminar flow minimizes viscous dissipation, resulting in the transition to elastic dissipation in this regime, as observed in Fig.~\ref{fig:f3}c.
    	
	This new regime of active nematics features fewer defects but still exhibits energy transfer across length scales. Much like defect chaos, it arises specifically through the ability to exchange distortions in the director field with deviations from $S=1$. In both regimes this process happens predominantly during a bend instability. In defect chaos, the instability results in topological defects that introduce discontinuities in the director field, thus transferring elastic energy to short wavelengths. However, in the new regime, the bend instability results in a bend wall with an extended profile, thus transferring energy to longer wavelengths. For this reason, we refer to this regime as ``bend wall chaos''.
    
    Bend wall chaos has many interesting features that are not yet well understood. For example the bend walls have a tendency to form spirals. This is because when two bend walls wrap around each other they generate a rotating flow. In addition, defects move primarily along bend walls, releasing stored elastic energy rather than through self propulsion as in typical defect chaos~\cite{giomi2014defect}. Positive defects often appear at the end of a bend wall, generating an active flow which counteracts their motion back along the bend wall; this can result in positive defects becoming stalled. This new regime is visually similar to arrested flow patterns observed in numerical studies on active nematics at the $S=1$ limit~\cite{lavi2024dynamical}, with the key distinction that in our simulations the patterns remain highly dynamic, with constant reforming of defects and bend walls, see SI Movie 2. Realizing bend wall chaos experimentally would require a system in which $\epsilon$ can be reduced independently from the elastic constant. \red{This is typically difficult but could be achieved by balancing two variables such as the stiffness and length of the constituent filaments of the active nematic.}
	
	In this manuscript we have demonstrated for the first time that energy transfer across length scales is an intrinsic feature of active nematic materials that allow for topological defects; this corresponds to the vast majority of experimentally realized active nematics. Furthermore, we show that energy can be transferred to larger or smaller length scales, depending on the the material parameters that affect the elastic energy.

    The active length determines the scale at which kinetic energy is injected into the system, which is translated into elastic energy through rotation of the director field. Energy transfer across scales arises primarily from the ability to exchange elastic energy between variations in the director field and the scalar order parameter, which are governed by the two components of the Landau De-Gennes free energy. 
     
    We show that during chaotic flows, elastic energy is constantly exchanged between the two components of the Landau De-Gennes free energy. \red{In defect chaos this happens predominantly during defect creation, resulting in energy transfer from large to small length scales.} When the energetic cost of topological defects is greatly increased, the active nematic instead forms long, dynamic bend walls which conversely transfer energy from small to large length scales, a regime we dub bend wall chaos. This is in contrast to traditional inertial turbulence in which dimensionality determines the direction of energy transfer. These results deepen our understanding of fundamental, yet unsolved, features of active fluids that continue to fascinate us both in their similarities and differences with classical fluids.
 

	\begin{acknowledgments}
		DJGP acknowledges funding from the Swiss National Science Foundation under starting grant TMSGI2 211367. J.I.-M., and F.S. acknowledge funding from MICIU/AEI/10.13039/501100011033 (Grant No. PID2022-137713NB-C21). The authors are indebted to the Brandeis University MRSEC Biosynthesis facility (supported by NSF MRSEC 2011846) for providing the tubulin to perform experiments. We thank Dom Corbett for insightful discussions.  
	\end{acknowledgments}
    \FloatBarrier
	\bibliography{bibfile}

\end{document}